\documentclass[aps,amsmath,superscriptaddress,longbibliography]{revtex4-1}

\usepackage{amsmath}
\usepackage{amssymb}
\usepackage[pdftex,bookmarks,colorlinks]{hyperref}
\pdfoutput=1
\usepackage[pdftex]{graphicx}
\usepackage{blindtext}
\usepackage{natbib}
\usepackage{upgreek}
\usepackage{subcaption}
\usepackage{calc}
\usepackage{amsmath}
\usepackage{amssymb}

\begin{document}

\title{Oblique drop impact onto a deep liquid pool}

\author{Marise V. Gielen} \email[]{m.v.gielen@utwente.nl} 
\author{Pascal Sleutel} \email[]{c.p.sleutel@utwente.nl}
\affiliation{Physics of Fluids Group, Faculty of Science and Technology, MESA+ Institute for Nanotechnology, University of Twente, P.O. Box 217, 7500 AE Enschede, The Netherlands.}
\author{Jos Benschop}
\affiliation{ASML The Netherlands B.V., De Run 6501, 5504 DR Veldhoven, The Netherlands}
\author{Michel Riepen}
\affiliation{ASML The Netherlands B.V., De Run 6501, 5504 DR Veldhoven, The Netherlands}
\author{Victoria Voronina}
\affiliation{ASML The Netherlands B.V., De Run 6501, 5504 DR Veldhoven, The Netherlands}
\author{Claas Willem Visser}
\author{Detlef Lohse}
\affiliation{Physics of Fluids Group, Faculty of Science and Technology, MESA+ Institute for Nanotechnology, University of Twente, P.O. Box 217, 7500 AE Enschede, The Netherlands.}
\author{Jacco H. Snoeijer}
\affiliation{Physics of Fluids Group, Faculty of Science and Technology, MESA+ Institute for Nanotechnology, University of Twente, P.O. Box 217, 7500 AE Enschede, The Netherlands.}
\affiliation{Mesoscopic Transport Phenomena, Eindhoven University of Technology, P.O. Box 513, 5300 MB Eindhoven, The Netherlands.}
\author{Michel Versluis}
\affiliation{Physics of Fluids Group, Faculty of Science and Technology, MESA+ Institute for Nanotechnology, University of Twente, P.O. Box 217, 7500 AE Enschede, The Netherlands.}
\author{Hanneke Gelderblom} \email[]{h.gelderblom@utwente.nl}
\affiliation{Physics of Fluids Group, Faculty of Science and Technology, MESA+ Institute for Nanotechnology, University of Twente, P.O. Box 217, 7500 AE Enschede, The Netherlands.}

\date{\today}

\begin{abstract}
Oblique impact of drops onto a solid or liquid surface is frequently observed in nature. Most studies on drop impact and splashing, however, focus on perpendicular impact. Here, we study oblique impact of 100-micrometer drops onto a deep liquid pool, where we quantify the splashing threshold, maximum cavity dimensions and cavity collapse by high-speed imaging above and below the water surface. Gravity can be neglected in these experiments. Three different impact regimes are identified: smooth deposition onto the pool, splashing in the direction of impact only, and splashing in all directions. We provide scaling arguments that delineate these regimes by accounting for the drop impact angle and Weber number.  The angle of the axis of the cavity created below the water surface follows the impact angle of the drop irrespectively of the Weber number, while the cavity depth and its displacement with respect to the impact position do depend on the Weber number. Weber number dependency of both the cavity depth and displacement is modeled using an energy argument. 
\end{abstract}

\pacs{47.55.D-, 79.20.Ds}
\maketitle
\section{INTRODUCTION}

In nature, oblique drop impact is ubiquitous. It is for example encountered in rain drop impact onto waves or puddles \cite{FLM:378372}, where it triggers air entrainment  \cite{Prosperetti} and aerosol generation \cite{Sellegri2006} that drive the global gas/liquid exchange. In agriculture, e.g. in pesticides crop spraying \cite{Peirce2016209}, oblique drop impact and the subsequent splashing and droplet rebound \cite{giletbush2012} is important. In industrial applications oblique impact occurs in e.g. metal spray deposition \cite{0965-0393-9-2-305} and direct fuel injection internal combustion engines \cite{Wang2012770}. In many of these applications, splashing is an unwanted side effect after impact; it decreases the deposition efficiency and may lead to a widespread contamination.
 
While drop impact and splashing is a topic widely studied (see e.g. \cite{ Sommerfeld1997, Josserand2016365, Thoroddsen2012560, Zhang2012402, FLM:339495, yarin2006,0951-7715-21-1-C01, zaleskisplash, FLM:134687, FLM:2459868, ray2012, FLM:9600982, PhysRevE.92.053022, thoroddsenslingshot, Tran2013}), surprisingly few papers (e.g. \cite{ikalo200697, ikalo2005661, Antonini2014, aboud2015}) deal with non-perpendicular or oblique impact. Drop splashing upon perpendicular impact has been studied on a solid substrate \cite{FLM:339495, yarin2006}, a thin liquid film \cite{0951-7715-21-1-C01, zaleskisplash, FLM:134687} and a deep liquid pool \cite{FLM:2459868, ray2012, FLM:9600982, PhysRevE.92.053022, thoroddsenslingshot, Tran2013}. In the latter case, a cavity is formed under the water surface \cite{Prosperetti}. Collapse of this cavity \cite{Bergmann2009381, PhysRevLett.100.084502, Michon2017} may lead to the pinch-off of small droplets \cite{FLM:378372, PhysRevLett.102.034502}. These droplets emerge from a high-speed microjet, and can reach velocities higher than the impact velocity \cite{Longuet-Higgins1995183}. The main parameter that governs splashing is the Weber number, $We=\frac{\rho D U^2}{\gamma}$ with drop diameter $D$, drop velocity $U$, liquid surface tension $\gamma$ = 0.072 N/m and liquid density $\rho$= 1000 kg/m$^3$. In addition, the splashing threshold depends on the Reynolds number, $Re=\frac{UD}{\nu}$, with $\nu$ the kinematic viscosity. A standard splashing threshold has the form of $We^{1/2}Re^{1/4}=K$ \cite{Sommerfeld1997, yarin2006} where K is a constant that depends on the surrounding pressure \cite{PhysRevLett.94.184505}, the thickness of the liquid layer, and the surface roughness when a drop impacts onto a rigid substrate.

To study oblique impact, experiments onto dry tilted plates \cite{Antonini2014, ikalo200697, ikalo2005661, aboud2015} and onto a moving plate \cite{1367-2630-11-6-063017} were performed. Bird et al. \cite{1367-2630-11-6-063017} showed that by moving the substrate, the velocity of the ejecta sheet and therefore the splashing threshold changes. Only a few studies \cite{Alghoul2011,Okawa2008,preChe,Liang2013,PhysRevE.92.053005,castrejnpita2016} focus on oblique impact onto wetted surfaces and thin liquid films. Gao and Li \cite{PhysRevE.92.053005} quantify a splashing threshold for impact onto a moving thin liquid film, but due to experimental complications such as the liquid inertia this approach is not feasible for a deep liquid pool. 

In numerical simulations, oblique impact onto a wetted substrate \cite{Cheng201511, Brambilla2013415, Ray20121386} shows a transition from crown splashing to single-sided splashing. In addition, crown formation and cavity formation \cite{Ray20121386} are studied. Due to the aforementioned experimental limitations these simulations lack validation. For oblique drop impact onto a wetted or dry surface, there is a clear influence of the impact angle on the splashing, but to our best knowledge no experiments of oblique drop impact onto a liquid pool have been reported in literature until now. 

Here, we present an experimental study of oblique drop impact onto a quiescent deep liquid pool. We provide details of  our experimental method in section \ref{exp0} and discuss a typical result for the impact and splashing phenomena in section \ref{exp5}. The angle of impact is varied systematically to quantify the splashing threshold and a model to explain these observations is presented in section \ref{exp2}. We also quantify the cavity formation and present a scaling law for the cavity formation in section \ref{exp3}. 

\section{EXPERIMENTAL METHODS} \label{exp0}

\begin{figure}
\centering
   \begin{subfigure}[b] {0.45\textwidth}
	\includegraphics[width=\textwidth]{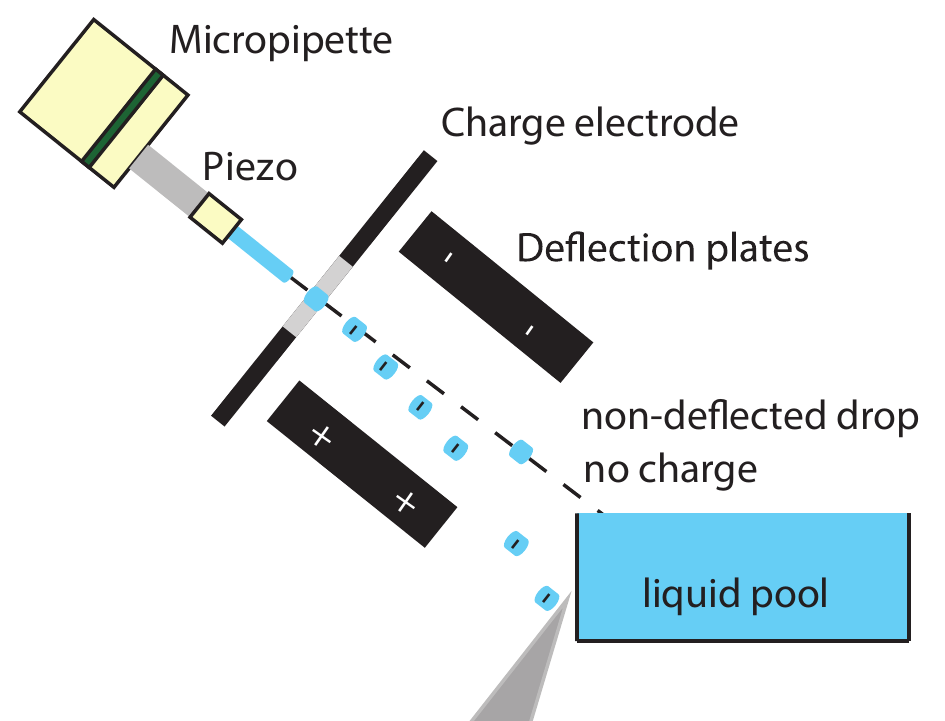}
       \caption{}
       \label{fig:setup}
    \end{subfigure}
    \begin{subfigure}[b] {0.45\textwidth}
	\includegraphics[width=\textwidth]{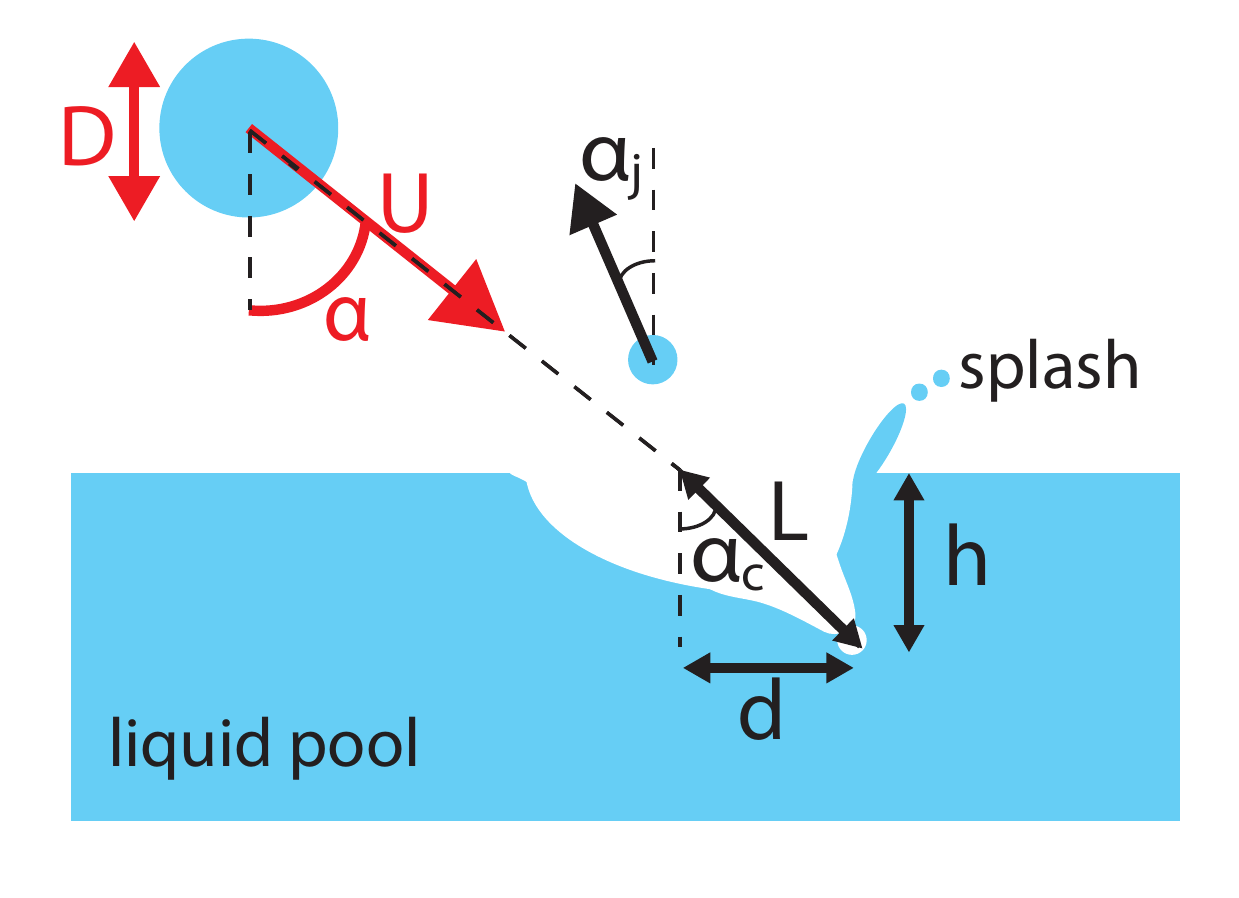} 
        \caption{}
        \label{fig:systematic}
    \end{subfigure}
\caption{a) Sketch of the experimental setup. A continuous stream of drops is created by piezoelectric-actuation of a microjet. A charging electrode charges the drops, which are subsequently deflected by the deflection plates. Every 10 milliseconds a single drop is left uncharged and can be separated from the stream. By tilting the micropipette, charging electrode and deflection plates, the impact angle $\alpha$ of the drops onto the pool can be varied. The pool has a depth of 10 mm, a width of 10 mm and a length of 100 mm. b) Drop diameter $D$, velocity $U$ and impact angle $\alpha$ are controlled by the experiment, and cavity displacement $d$ and height $h$ are measured, which results in cavity angle $\alpha_c$ and characteristic length of cavity $L$. Next to that, the angle of the jetted droplet out of the cavity $\alpha_j$ is measured. The shape of the splash and the collapse of the cavity are recorded by two separate high-speed recordings above and below the water surface, respectively.}
\label{fig:schematic}
\end{figure}

To study oblique drop impact for a wide range of Weber numbers and impact angles onto a deep liquid pool, two steps are important: (i) creation of a single drop and (ii) rotation of the drop generator to obtain oblique impact. To create single drops a method previously described in \cite{C4SM02474E} is used, which we adopt for oblique impact. For clarity, we briefly describe this method to isolate single drops from a stream of drops here, which is schematically drawn in Fig.~\ref{fig:setup}. 

Drops of size of 115~$\pm$~15~$\mu$m are generated by pumping (Shimadzu LC-20AD HPLC pump) demineralized water ($<$0.1\% of ammonia added for conductivity, negligible effects on surface tension) through a micropipette (Microdrop AD K-501). The continuous jet breaks up into drops by applying a piezo acoustic pressure on the jet, which transforms the jet into a stream of monodisperse drops with equal velocity. Velocities ranging from 6 up to 25 m/s can be achieved resulting in Weber numbers between 40 and 1056. Here, the lower bound is set by the minimum velocity required to create a train of drops and the upper bound by the maximal flow rate of the apparatus. The drop train is directed through a ring-shaped charging electrode, where a periodic high-voltage pulse charges all passing drops except one every 10 milliseconds. Next, the drops pass another region with a high electric field ($E \geq 100$ kV/m) in between two deflection plates. The electric field separates the charged drops from the uncharged ones, and allows for the uncharged drops to continue straight and impact onto the pool. The pool consists of a glass container that is filled to the top with water to minimize the influence of a disturbing meniscus at the container wall that may limit the optical imaging quality. The charged drops are caught further on their path and then disposed. To create an oblique impact, the micropipette, charge electrode and deflection plates are tilted. In this way, impact angles up to 80$^\circ$ from perpendicular can be obtained. This method allows for the creation of oblique drop impact with a wide range of velocities and a large enough drop spacing to prevent disturbances of the pool due to earlier impact events.

Figure ~\ref{fig:systematic} shows a sketch of drop velocity $U$ and impact angle $\alpha$. These parameters are varied systematically, while keeping the drop diameter $D$ constant.By doing so, we effectively change the ratio between the perpendicular impact velocity $U_{\bot} = U\cos\alpha$ and the  parallel impact velocity $U_{\parallel} = U\sin\alpha$. Note that $U_{\bot}$ and $U_{\parallel}$ cannot be varied completely independently since a minimum total velocity $U$ is required to create a train of drops.

Drop impact is recorded in a side view by high-speed imaging (Photron SA-X2, operated at frame rates up to 100 kfps, average pixel resolution of 3.2 $\upmu$m) and backlight illumination (Olympus ILP-2). From these recordings, we extract $D$, $U$ and $\alpha$ using a customized image-processing analysis. The impact velocity is measured from the penultimate frames just before impact to minimize the influence of air drag. The error in these measurements is estimated to be 2 $\mu$m for $D$ and 0.2 m/s for $U$, both based on the camera resolution. For a typical measured velocity of $U$ = 15 m/s, these errors result in an error in $We$ of about 15 and in $\alpha$ of about $2^\circ$.

The dynamics of the cavity is characterized simultaneously with the impact event by collecting two separate recordings for each impact condition, one above and one below the water surface, in order to compensate for the different optical focal depth. From the measurements above the water surface we characterize the direction of the splash. From the cavity recordings, we estimate the maximum cavity displacement $d$, maximum height $h$ and define the cavity angle $\alpha_c$ by $\tan\alpha_c = \left(\frac{d}{h}\right)$. These maximum dimensions are not always clearly pronounced. We therefore obtain $d$ and $h$, and hence $\alpha_c$, from an average of six measurements above and below the water surface for each impact condition. The typical errors obtained are a few $\upmu$m in $d$ and $h$ and a few degrees in $\alpha_c$. In total, to quantify the cavity dimensions and collapse below the water surface and the splashing threshold above the water surface, we analyzed 1147 individual drop impacts. 

\section{RESULTS AND INTERPRETATION} \label{exp1}

\subsection{Typical features of oblique drop impact} \label{exp5}

\begin{figure*}
\centering
\includegraphics[width=12cm]{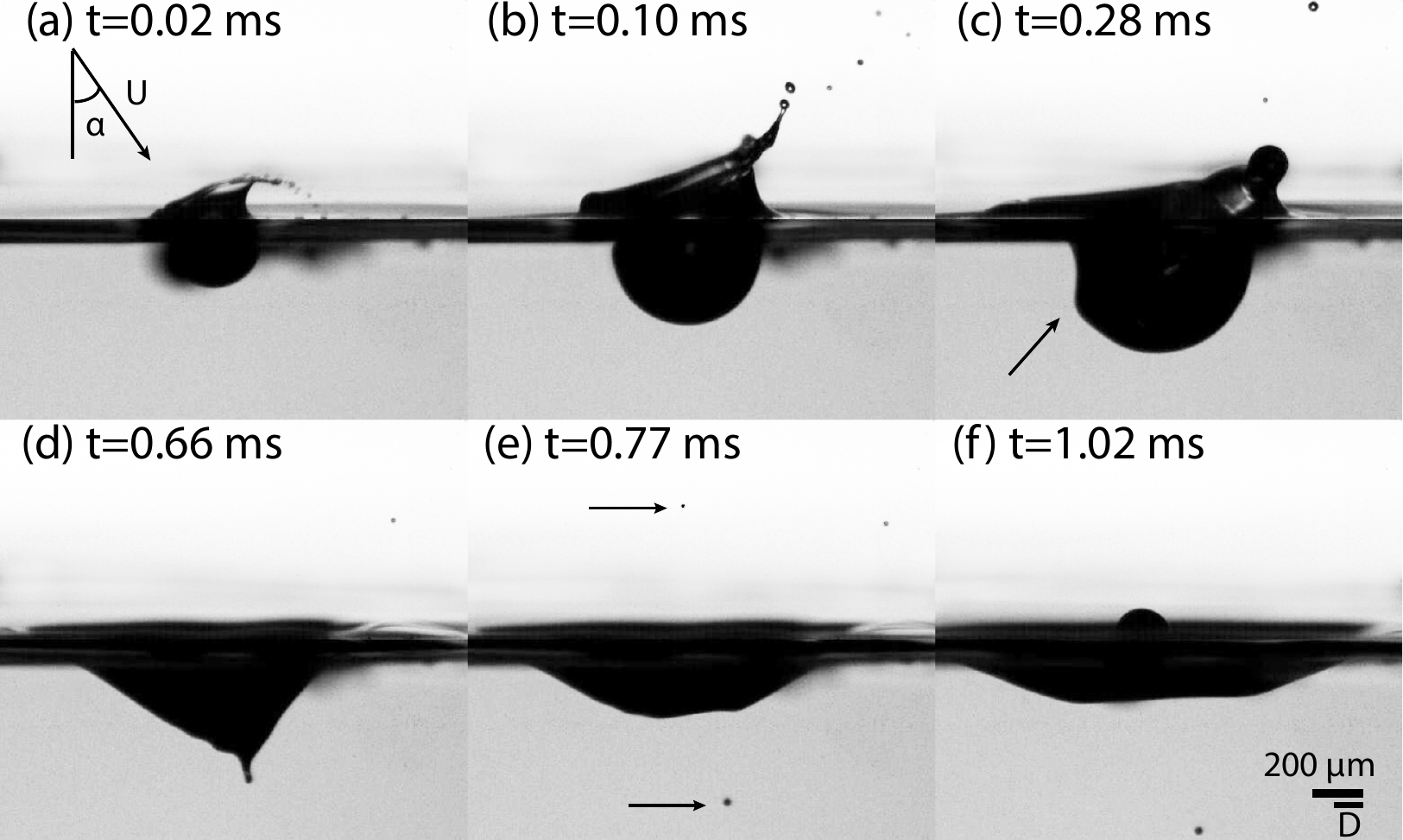}
\caption{Time series of a drop impacting onto the pool with a Weber number $We$ = 674 and impact angle $\alpha$ = $28^\circ$, where t=0 marks the moment the drop makes first contact with the pool. In this caption time $t$ is non-dimensionalized to $t^* = \frac{t}{t_{imp}} = \frac{tU}{D}$ while in the figure the dimensional time is given. (a) $t^*$ = 3.6. The impacting drop causes a splash that fragments into droplets. Here, the splash is formed on a single side only. (b) $t^*$ = 18. A crown develops and remains more pronounced on one side. The cavity has a hemispherical form and is growing. (c) $t^*$ = 51. The splashing is finished and the cavity still grows. An arrow points to the wave crest, which travels downwards along the cavity surface, see \cite{PhysRevLett.100.084502}. The oblique impact leads to an asymmetry of the cavity. (d) $t^*$ = 119. The capillary wave focuses off-center at the bottom of the cavity. (e) $t^*$ = 139. Collapse of the wave leads to the entrainment of a bubble and retraction of the jet leads to the ejection of tiny droplets during closure of the cavity. The arrows indicate the position of the entrained bubble and the droplets. (f) $t^*$ = 184. A Worthington jet is formed during closure of the cavity~\cite{worthington}.}
\label{fig:Series}
\end{figure*}

Figure \ref{fig:Series} shows the time series of a typical experiment with $We$ = 674 and $\alpha$ = $28^\circ$.  Qualitatively the different stages of oblique impact are comparable to perpendicular impact, but with some important qualitative differences, which we will now discuss.

Panel (a) shows a snapshot 0.02 ms after the drop has impacted. The formation of a hemispherical cavity has started below the water surface and an asymmetric splash ejecting small droplets is visible above the water surface. At $t$ = 0.1 ms, see panel (b), both the crown and the cavity are growing. In contrast to perpendicular impact the droplets only detach on a single side. Just below the water surface, a wave crest is observed~\cite{PhysRevLett.100.084502}. At $t$ = 0.28 ms (panel (c)) this wave crest is highlighted by an arrow while it travels downward along the cavity wall. Here one can clearly observe that the cavity is asymmetric. At this point in time, the cavity is still growing while surface tension causes the crown to retract and the rim to grow thicker. Panel (d) ($t$ = 0.66ms) shows the cavity at its deepest point, where the maximum cavity depth and displacement are reached. Then, the closure of the cavity starts and the capillary wave crest collides at the bottom of the cavity. During the collision of these waves an air bubble can be entrapped \cite{Prosperetti}. The tip at the bottom of the cavity in (d) marks this air bubble just before pinch-off. The tip will retract quickly due to the high local curvature. The series of subsequent image frames suggest that the jetted droplet, which is visible above water surface in panel (e) at $t$ = 0.77 ms, originates from the rapid tip retraction. The droplets jetted out of the cavity reach velocities of the same order as the main drop impact velocity. Below the water surface (panel (e)) a tiny bubble is clearly visible. At $t$ = 1.02 ms (panel (f)) the cavity has completely collapsed, marked by the emergence of a thick Worthington jet \cite{worthington}. 

\subsection{Splashing threshold} \label{exp2}
\begin{figure}
\centering
   \begin{subfigure}[c]{1\columnwidth}
        \includegraphics[width=\textwidth]{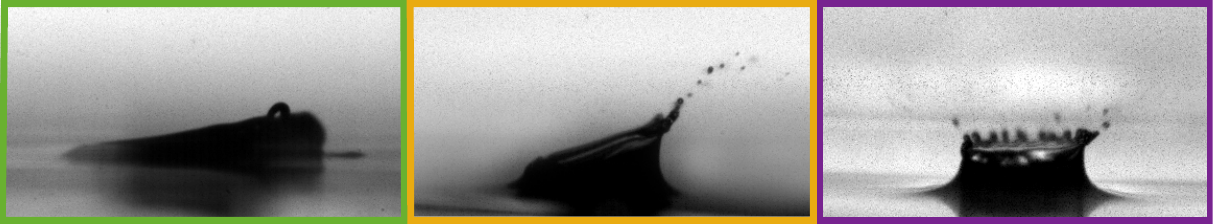}
        \caption{}
        \label{fig:splashimage}
    \end{subfigure}
 \\
    \begin{subfigure}[c]{1\columnwidth}
           \includegraphics[width=10cm]{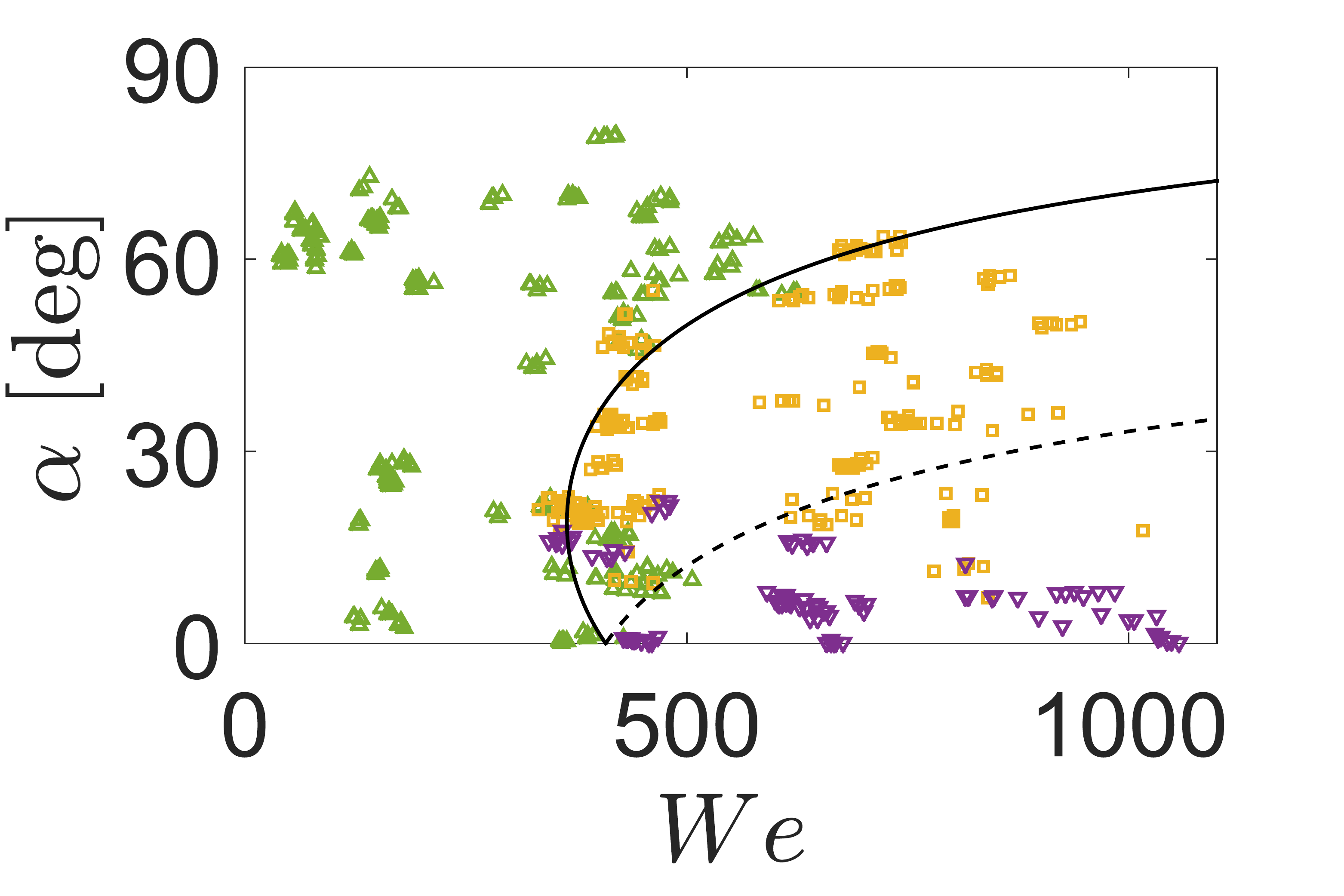}
       \caption{}
       \label{fig:phase}
    \end{subfigure}
\caption{a) Three types of impact behavior are observed above the water surface. From left to right: deposition (green), single-sided splashing (yellow) and omni-directional splashing (purple). b) A phase diagram of the impact behavior as a function of the Weber number $We$ and impact angle $\alpha$. The color codes correspond to Fig.~a): upward green triangles represent deposition, yellow squares represent single-sided splashing and downward purple triangles represent omni-directional splashing. The solid and the dashed line are derived from Eqn.~(\ref{eq:Vc}) with $c=0.44$, where the solid line represents the splashing threshold from deposition to single-sided splashing (Eqn.~(\ref{eq:Vc}) with plus sign) and the dashed line indicates the splashing threshold from single-sided splashing to omni-directional splashing (Eqn.~(\ref{eq:Vc}) with minus sign).}
\label{fig:threshold}
\end{figure}

After drop impact three different phenomena can be observed above the water surface, as illustrated in  Fig.~\ref{fig:splashimage}. First, the drop can smoothly coalesce with the liquid pool. This behavior is identified as deposition. Second, a crown can be formed, which destabilizes on a single side of the drop and results in the ejection of satellite droplets at the tips of the crown on this side of the drop. The direction of splashing corresponds to the direction of impact. This behavior is denoted as single-sided splashing. The final phenomenon observed is the ejection of small satellite droplets from the crown in all directions, which is indicated as omni-directional splashing. We quantify these phenomena in a phase diagram of $\alpha$ and $We$ illustrated by three distinct regions, see Fig.~\ref{fig:phase}. For a single value of $\alpha$ the impact behavior can change from deposition to single-sided splashing to omni-directional splashing for increasing $We$. Similarly, for a single value of $We$ the impact behavior can change from omni-directional splashing to single-sided splashing to deposition for increasing $\alpha$.

\begin{figure}[!]
\centering
   \begin{subfigure}[b] {0.45\textwidth}
	\includegraphics[width=\textwidth]{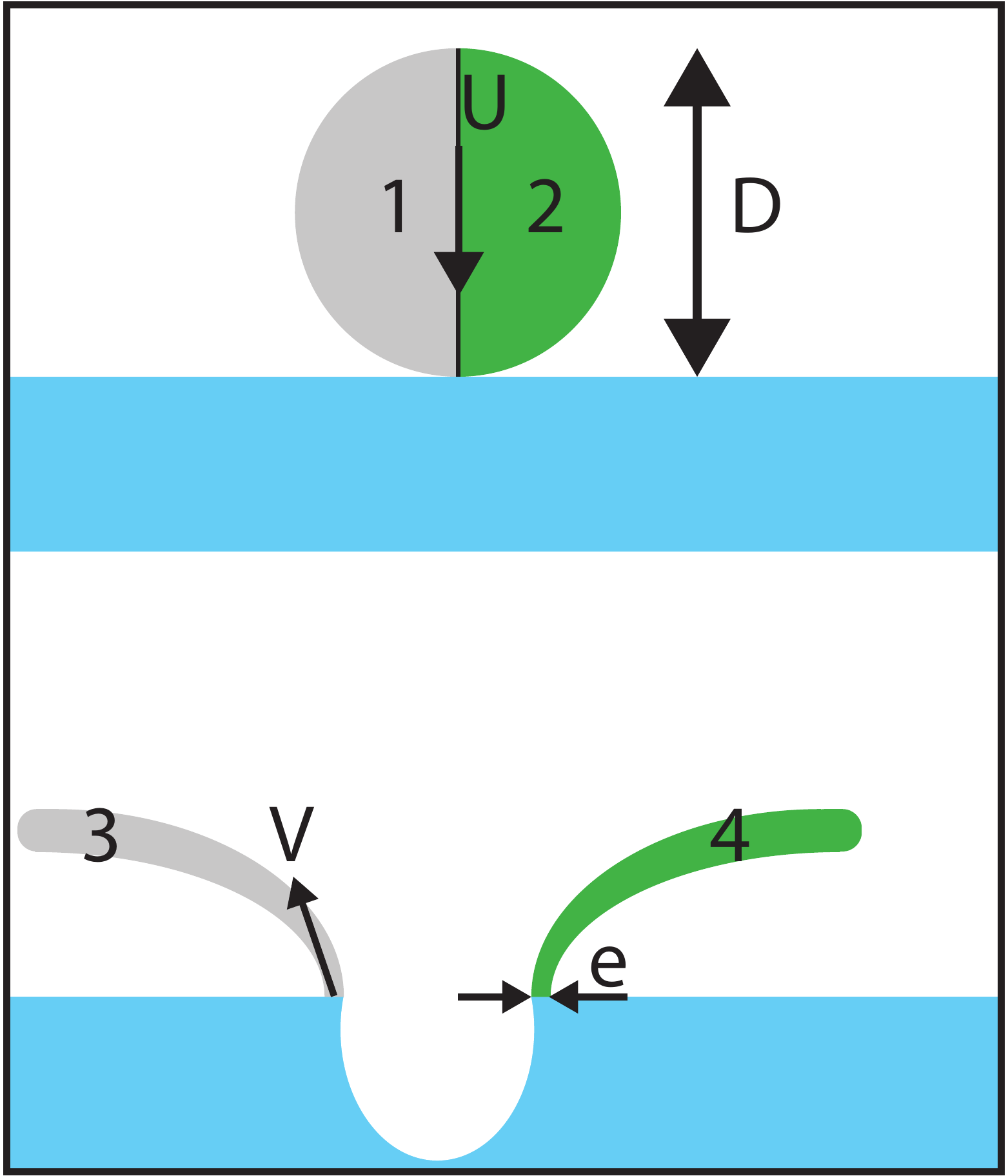} 
	\caption{} 
	\label{fig:fullimp}
    \end{subfigure}
    \begin{subfigure}[b] {0.45\textwidth}
	\includegraphics[width=\textwidth]{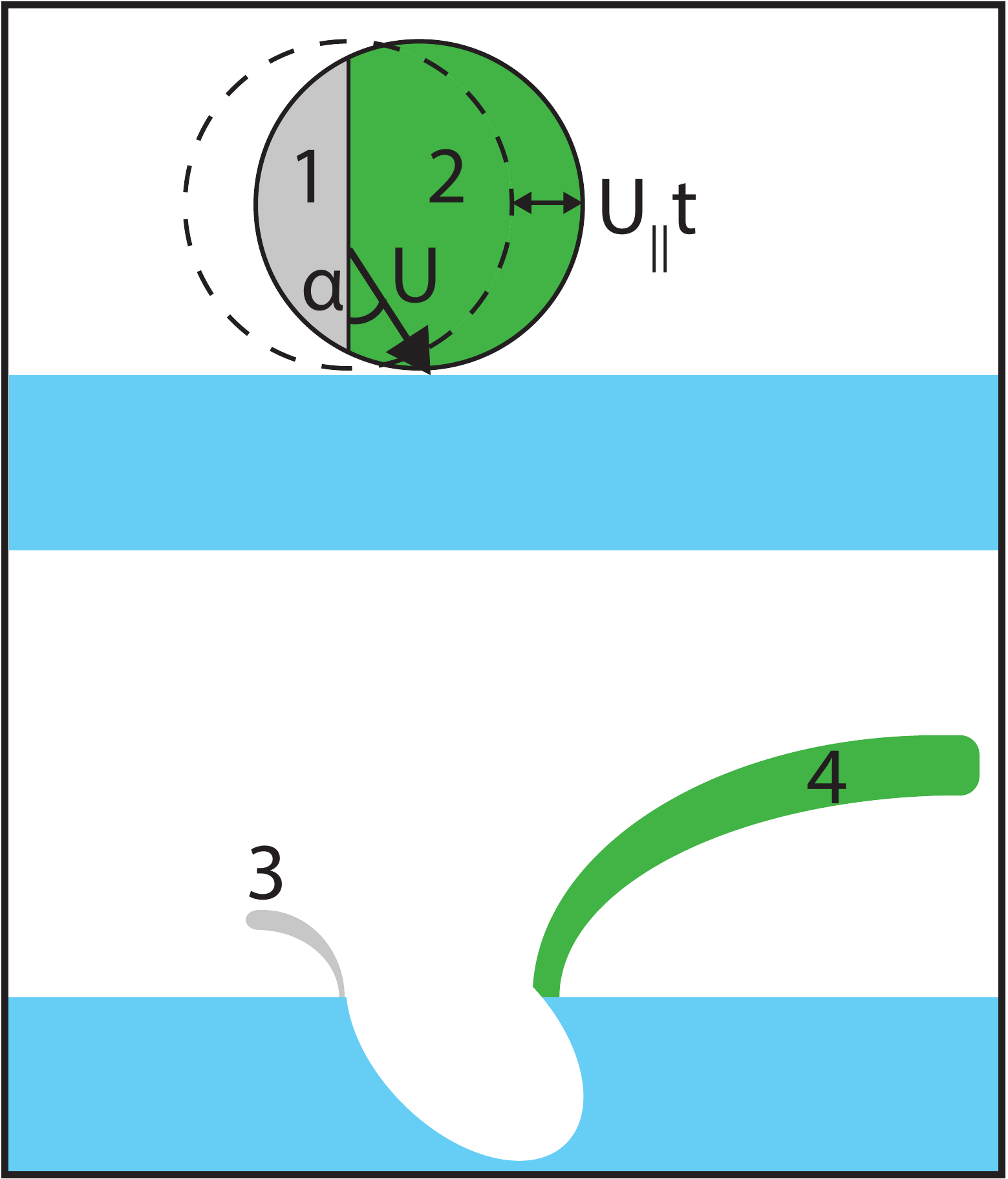} 
	\caption{} 
	\label{fig:shift}
    \end{subfigure}
\caption{Schematic illustration of the mass distribution in the crown. Volume 1 indicates the part of the original drop that flows into crown volume 3. Volume 2 represents the part of the original drop that flows into crown volume 4. Volume 1 and 3, and 2 and 4 are identical due to mass conservation. a) For perpendicular drop impact the volume of the original drop is distributed evenly over the crown, such that volume 1 and 2, and thus 3 and 4 are identical. b) For oblique impact, $U_{\parallel}$ will cause a shift (solid line) in drop position with respect to the perpendicular impact (dashed line). This shift reduces volume 1 and increases volume 2 by an amount of $\pm \rho DU_{\bot}U_{\parallel}t$ (see text). As a result, the crown becomes asymmetric.}
\label{fig:fis}
\end{figure}

We now provide a scaling argument to explain these observations. 
To find the splashing threshold for oblique impact, we aim to describe the velocity of the crown in terms of $U$, $D$, $\alpha$ and $We$. We start by considering perpendicular (i.e. vertical) impact. A schematic view of the drop during splashing is shown in Fig.~\ref{fig:fullimp}. In this case, the flow into the crown is distributed symmetrically. Once the entire drop has impacted, we can assume that the entire drop volume is proportional to the volume of the crown, which gives (per unit time)~\cite{zaleskisplash}
\begin{equation}
D^2U\sim eDV,  \label{eq:mass}
\end{equation}
where $e$ is the thickness of the crown at its origin. In line with a model for splashing of the fast ejecta onto a thin liquid film~\cite{zaleskisplash}, we assume that in order to obtain splashing upon impact onto a liquid pool the velocity of the crown $V$ has to be larger than the Taylor Culick velocity $V_{TC}\sim\sqrt{\frac{\gamma}{\rho e}}$. Thoroddsen \cite{thoroddsenslingshot} showed experimentally that $e\sim\sqrt{\nu t}$, consistent with dimensional analysis. Using $t \sim \frac{D}{U}$ in the expression for $e$ and~(\ref{eq:mass}) to express V we find the splashing criterion
\begin{equation}
\frac{V}{V_{TC}}  \sim We^{1/2}Re^{1/4} > K, \label{eq:Vcl}
\end{equation}•
where K is the critical number for splashing~\cite{Stow419, Sommerfeld1997, Josserand2016365}.
• 
The numerical value of $K$ depends on specific experimental conditions, such as the flow profile inside the crown~\cite{zaleskisplash}. Previous studies reported $K \approx 54$ for a rigid surface~\cite{Josserand2016365}, $K \approx 160$ for a thin liquid film~\cite{0951-7715-21-1-C01} and $K \approx 90$ for a deep liquid pool~\cite{PhysRevE.92.053022}. For our experiments we determine for perpendicular impact ($\alpha \simeq 0^{\circ}$) a critical Weber number for splashing of about $We_s \approx 400$ which leads to $K\approx 130$.

For oblique impact the mass is no longer symmetrically distributed over the crown: the mass flow into the crown increases on one side of the drop and decreases on the other side of the drop (Fig.~\ref{fig:shift}). This unequal mass distribution originates from the contribution of the tangential component of the impact velocity $U_{\parallel}$. On one side of the drop, $U_{\parallel}$ leads to an increased mass flow while on the other side the mass flow is decreased. The added/reduced mass flow rate into the crown scales as $\pm \rho DU_{\bot}U_{\parallel}t$ respectively, where $U_{\bot}$ sets the speed with which the drop is moving down into the pool. To account for this additional mass flow, we rewrite~(\ref{eq:mass}) as
\begin{equation}
 D^2U_{\bot}\pm cDU_{\bot}U_{\parallel}t\sim eDV, \label{eq:massob}
\end{equation}•
where c is a fit parameter that accounts for the exact amount of mass that is redistributed over the crown.
The (critical) velocity for crown splashing now depends both on $U_{\bot}$ and $U_{\parallel}$. From~(\ref{eq:massob}) it follows that $U_{\bot}\left[ 1\pm c \frac{U_{\parallel}t}{D}\right]\sim\frac{Ve}{D}$ and hence the splashing criterion for oblique impact reads, with $t\sim\frac{D}{U_{\bot}}$,
\begin{equation}
\frac{V}{V_{TC}}  \sim We^{1/2}Re^{1/4}(\cos{\alpha_s})^{5/4}\left[1\pm c\tan{\alpha_s}\right]  > K. \label{eq:Vc}
\end{equation}•
In~(\ref{eq:Vc}), we use $K=130$ (as determined for perpendicular impact) to find the splashing threshold as a function of $We$ and $\alpha_s$. These considerations show that there are two transitions in the phase space of Fig~\ref{fig:phase}: (i) A transition from deposition to single-sided splashing when $V_{TC}$ is reached on a single side of the drop (solid line). (ii) A transition from single-sided splashing to omni-directional splashing when $V_{TC}$ is reached for splashing in all directions (dashed line). 

In Fig.~\ref{fig:phase} we plot these transitions and we find qualitative agreement with the experimental data using $c=$ 0.44.The model also quantitatively captures the transition from deposition to single-sided splashing, while the predicted transition from single-sided to omni-directional splashing shows a stronger deviation from experiment. Moreover, we experimentally observe a zone where all three impact behaviors overlap for $We$ around 400 and $\alpha < 20^\circ$, which is absent in the model. In this transition region the three impact behaviors are close together. As a consequence, this region is very sensitive to small experimental variations in $We$ and $\alpha$. In addition, smaller droplets ejected in a splash might not be observable because of the camera resolution, which further complicates the judgment on the impact behavior. More importantly, three-dimensional effects that occur outside the observation plane make it difficult to discriminate between single-sided splashing and omni-directional splashing in a discrete manner: the transition is not sharp but gradual. Therefore the dashed line in Fig.~\ref{fig:phase} is far from perfect in separating between single-sided and omni-directional splashing.

\subsection{Cavity formation} \label{exp3}

\begin{figure}
\centering
\includegraphics[width=8cm]{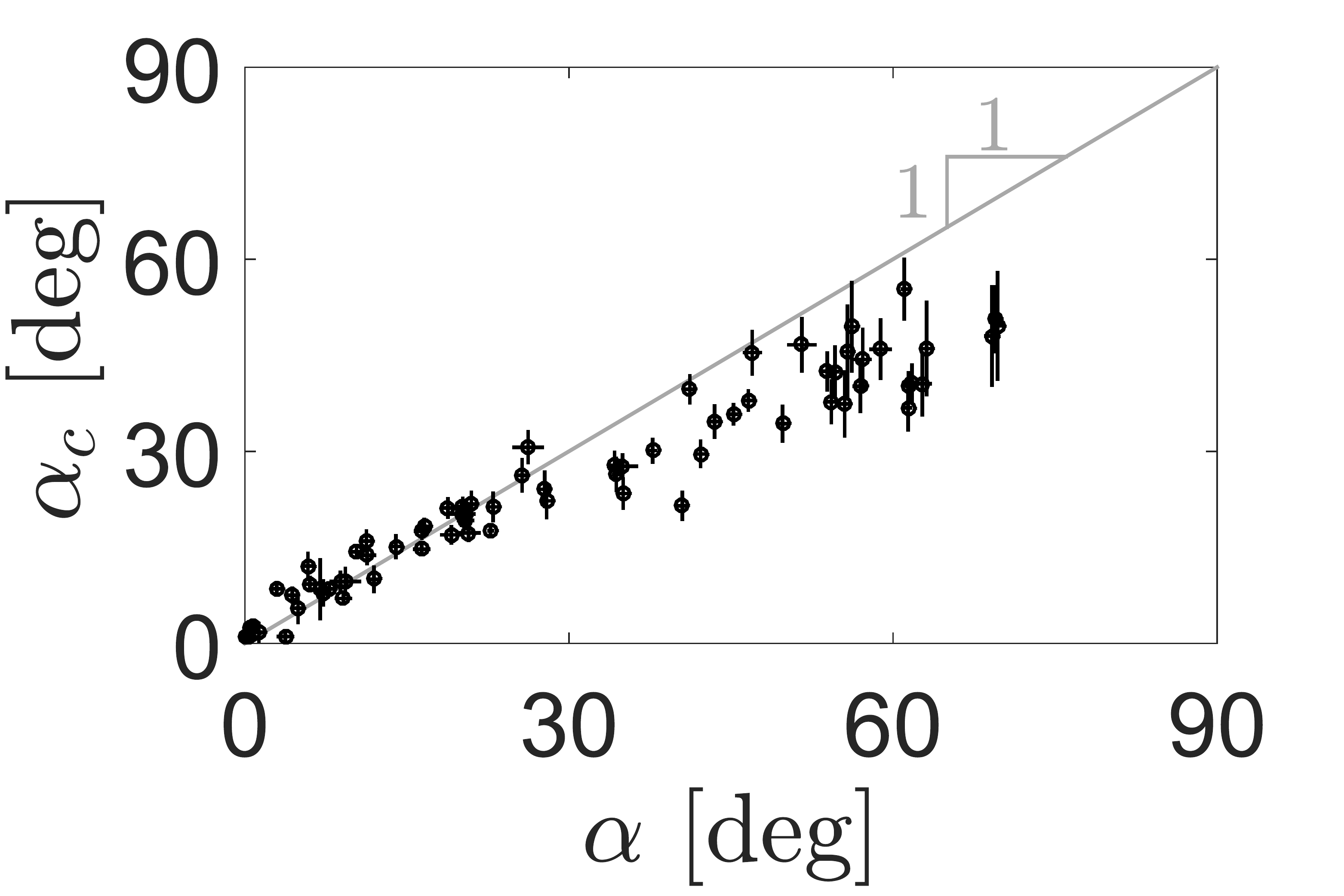}
\caption{Cavity angle $\alpha_c$ as a function of the impact angle $\alpha$. The solid line has slope unity. Each data point is the average value of six separate experiments, where the error bars indicate the propagated error in $\alpha_c$ corresponding to the measurement error in $d$ and $h$.}
\label{fig:alphaic}
\end{figure}

From our recordings we extract the angle $\alpha_c$ (as defined in fig~\ref{fig:systematic}) and the dimensions ($h$ and $d$) of the cavity. In Fig.~\ref{fig:alphaic} cavity angle $\alpha_c$ is plotted as a function of impact angle $\alpha$. For small impact angles the cavity angle equals the impact angle, which is expected since the momentum of the drop is transferred to the cavity in the direction of impact. However, for impact angles larger than $\alpha\gtrsim30^\circ$, $\alpha_c$ gets increasingly smaller as compared to $\alpha$. The measurement error of $\alpha_c$ does not explain the observed decrease. We analyzed the cavity angle as a function of $We$, but found no significant dependency (data not shown). A possible explanation for the deviation in proportionality of $\alpha_c$ with $\alpha$ for larger impact angles could come from additional wave drag~\cite{LeMerrer13092011} for drops moving almost parallel to the surface (i.e. drops impacting with a large $\alpha$). For larger impact angles, part of the impact energy will be dissipated into the build-up of waves in the direction tangential to the surface, which reduces the cavity angle.

\begin{figure}[!]
\centering
	\includegraphics[width=0.5\textwidth]{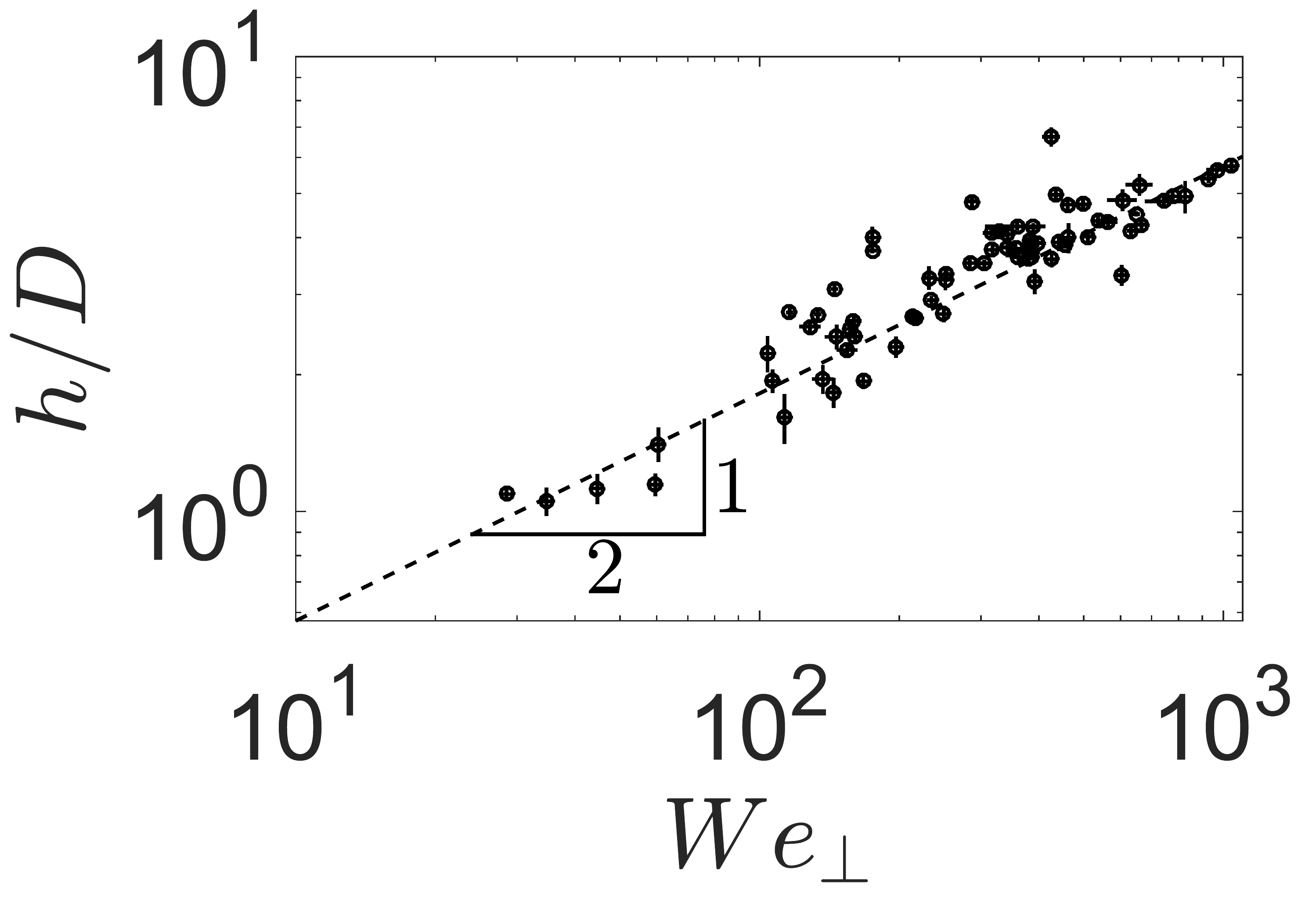}
	\label{fig:deplog}
\caption{Double logarithmic plot of the maximum cavity depth $h$ as a function of the Weber number, based on the perpendicular impact velocity $We_{\bot}$. The dashed line corresponds to Eqn.~(\ref{eq:hD}) and a prefactor of 0.18. Each data point is the average value of six separate experiments with corresponding error bars.}
\label{fig:depth}
\end{figure}
\begin{figure}[!]
\centering
	\includegraphics[width=0.5\textwidth]{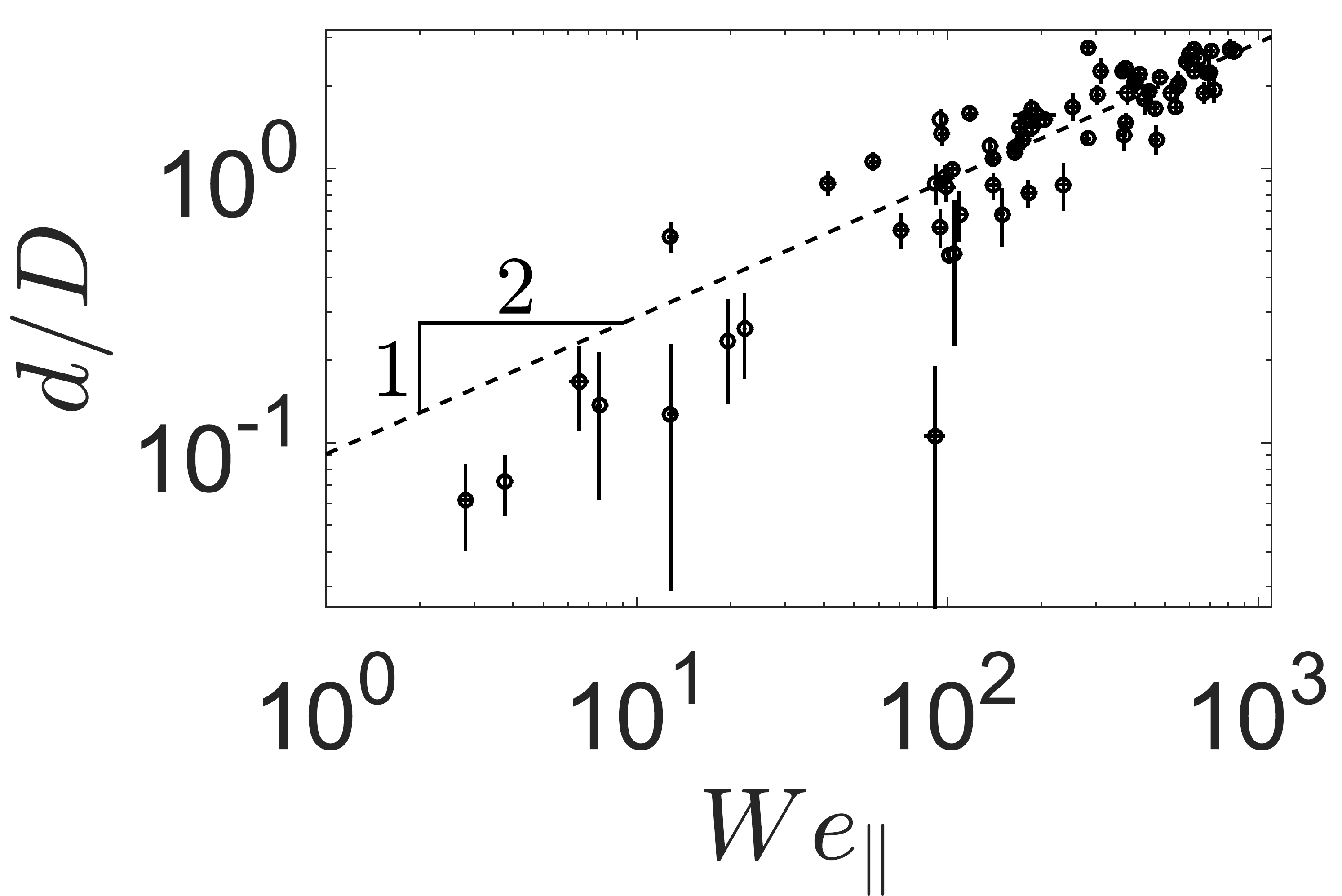} 
	\label{fig:displog}
\caption{Double logarithmic plot of the maximum horizontal cavity displacement $d$ as a function of the Weber number based on the tangential impact velocity  $We_{\parallel}$. The dashed line corresponds to Eqn.~(\ref{eq:dD}) and a prefactor of 0.09. Each data point is the average value of six separate experiments with corresponding error bars.}
\label{fig:displacement}
\end{figure}•
We now present data on the maximum cavity depth and maximum displacement. In Fig.~\ref{fig:depth} and Fig.~\ref{fig:displacement} the maximum cavity depth $h$ and displacement $d$ are plotted as a function of $We$ based on the perpendicular impact velocity $We_{\bot} = \frac{\rho D U^2_{\bot}}{\gamma}$ and the tangential velocity $We_{\parallel} = \frac{\rho D U^2_{\parallel}}{\gamma}$, respectively. The data suggests  $h\sim We_{\bot}^{1/2}$ while $d\sim We_{\parallel}^{1/2}$.
These scalings can be explained from energy conservation. We consider that the kinetic energy of the impacting drop is proportional to the additional surface energy to create the cavity
\begin{equation}
\rho D^3 U^2 \sim \gamma L^2, \label{eq:energy}
\end{equation}
where $L$ is the characteristic length of the cavity, as shown in Fig.~\ref{fig:systematic}. For oblique impact experiments, $L$ depends on both $h$ and $d$ since the maximum depth of the cavity is displaced by $d$ as a consequence of the tangential component of the impact velocity, resulting in $L\sim \sqrt{h^2+d^2}\sim h\sqrt{1+\tan^2{\alpha}}$. Here we have assumed $\tan\alpha=\frac{d}{h}$, which will introduce a small error for $\alpha\gtrsim30^\circ$, as seen in Fig.~\ref{fig:alphaic}. 
From~(\ref{eq:energy}) we find $\frac{L}{D} \sim We^{1/2}$. Using the geometrical relation for L we find $\frac{h}{D} \sim We^{1/2} \cos\alpha$, and hence
\begin{equation}
\frac{h}{D} \sim We^{1/2}_{\bot}. \label{eq:hD}
\end{equation}
Fig.~\ref{fig:depth} shows scaling law (\ref{eq:hD}) together with the experimentally determined maximum cavity depth, where we used a prefactor of 0.18. For $We_{\bot} > 100$, we observe a somewhat larger spread in $h$. This spread could be due to capillary waves propagating over the cavity surface, which distort the measurement of $h$. For the maximum cavity displacement we find $\frac{d}{D} \sim \frac{h}{D} \tan\alpha \sim We^{1/2} \sin\alpha$, which gives
\begin{equation}
\frac{d}{D} \sim We^{1/2}_{\parallel}. \label{eq:dD}
\end{equation}
In Fig.~\ref{fig:displacement} scaling~(\ref{eq:dD}) is plotted with a prefactor 0.09 together with the experimentally determined maximum cavity displacement. For $We_{\parallel} \lesssim$ 100 we observe a deviation between the experimental data and the proposed scaling. For these small Weber numbers the uncertainty in $d$ (of a few micron) becomes of the order of the value of $d$ that is actually measured. As explained in section~\ref{exp0}, the perpendicular and parallel impact velocity in our experiment cannot be varied completely independently. Indeed, to achieve the measurements at small $U_{\parallel}$, and hence small $We_{\parallel}$, we had to use a large $U_{\bot}$ to create a drop train. However, a large $U_{\bot}$ at small $U_{\parallel}$ implies a small $\alpha$ and hence a small $d$, which is hard to measure accurately. In addition, the measure for the cavity displacement is not as well defined as that for the cavity depth: when the cavity becomes less sharp, several values for $d$ could correspond to the same $h$, which leads to a further spread in the data in Fig.~\ref{fig:displacement}.


\section{Discussion on cavity collapse}  \label{exp4}

\begin{figure}[!]
\centering
    \begin{subfigure}[c] {0.45\textwidth}
	\includegraphics[width=\textwidth]{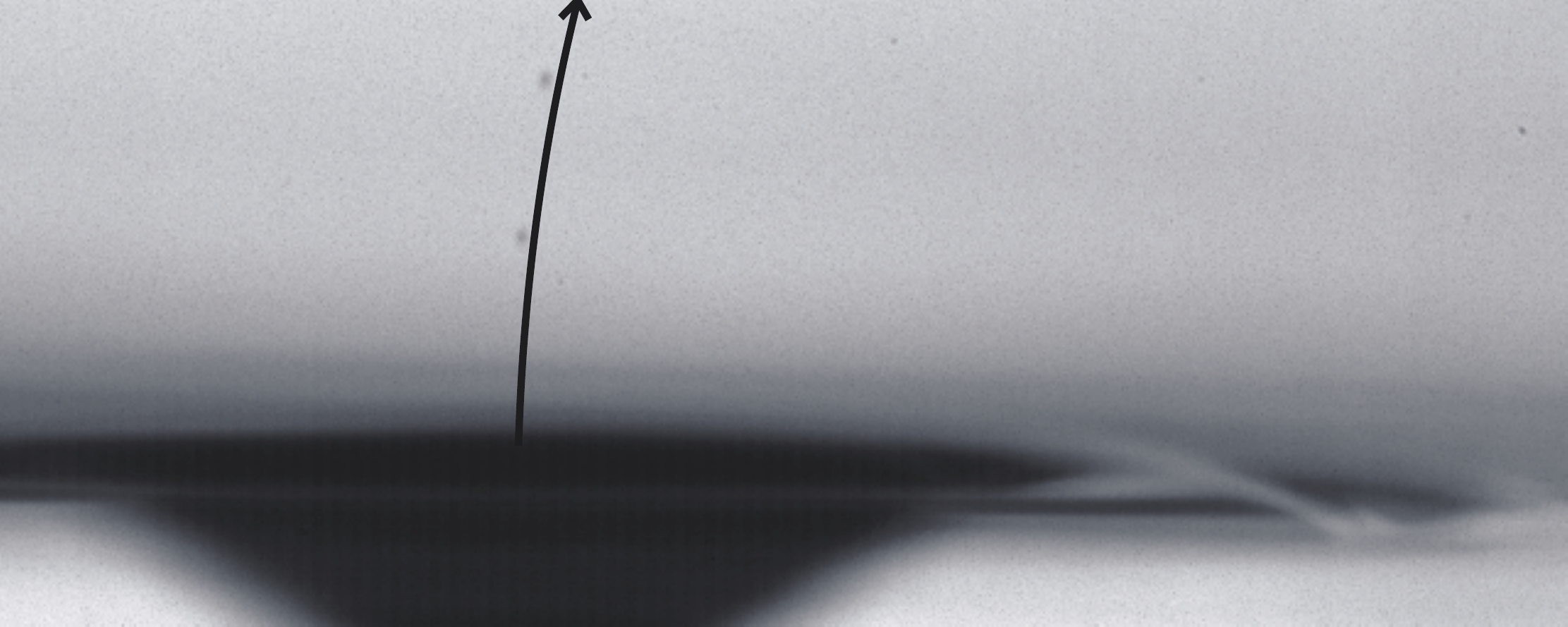} 
        \caption{}
        \label{fig:trajectory}
    \end{subfigure}
      \begin{subfigure}[c] {0.45\textwidth}
	\includegraphics[width=\textwidth]{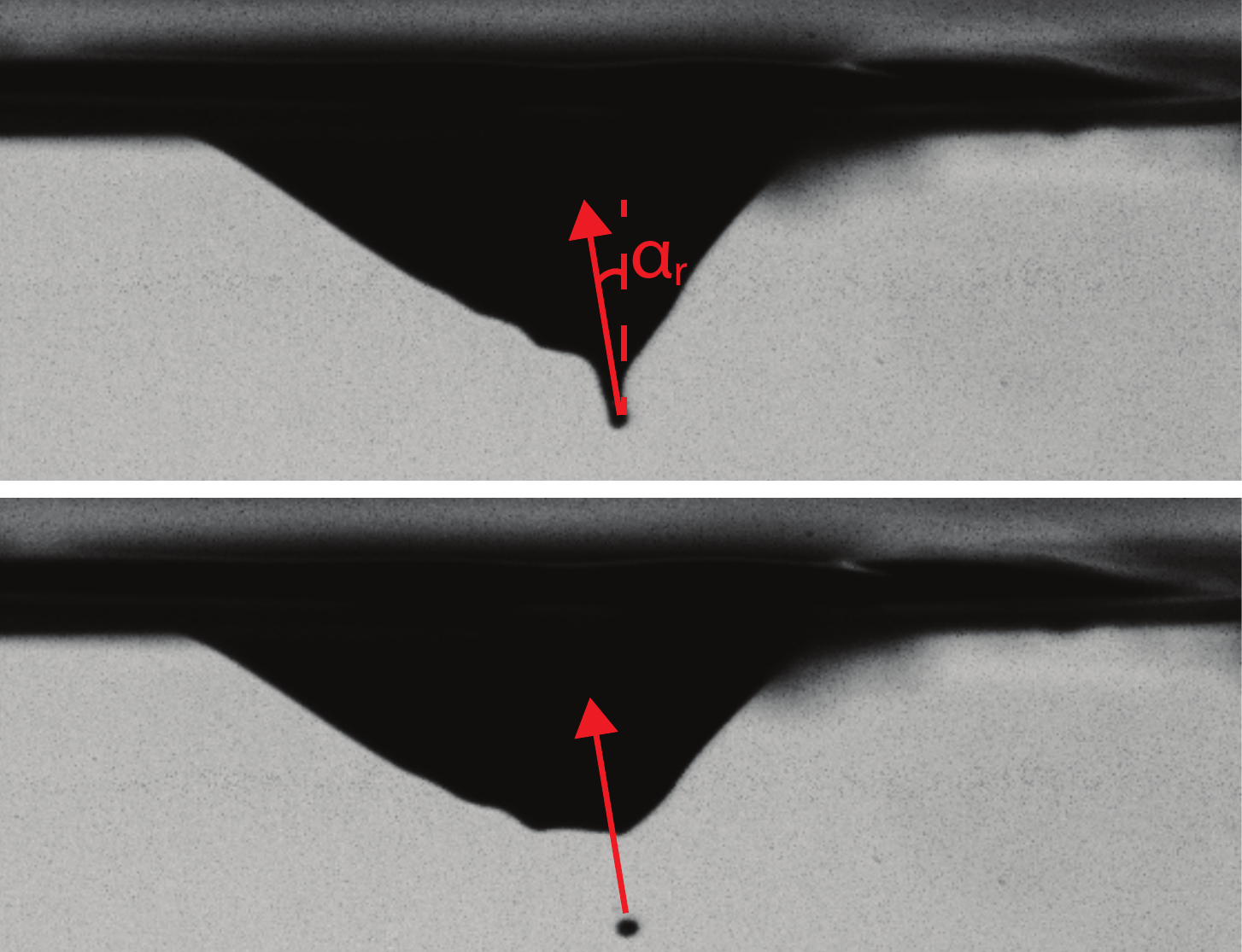} 
       \caption{}
       \label{fig:alphar}
    \end{subfigure}
\caption{a) The trajectory of two jetted droplets for We~$= 371$ and $\alpha = 0.3^\circ$ shows a deflection when the droplets leave the cavity. This deflection is indicated by the arrow. b) Snapshot of the cavity at its maximum dimensions (top), and approximately 14 microseconds later (bottom) when the collapse and tip retraction have started. The red arrows indicate the direction in which the cavity starts to collapse, which is used as a measure for the direction of the jetted droplet out of the cavity. }
\label{fig:schematic}
\end{figure}

The collapse of a cavity is sometimes accompanied by the jetting of a droplet out of the cavity, see Fig.~\ref{fig:trajectory}. In this subsection we aim to quantify the direction of the jetting behavior. Figure~\ref{fig:systematic} depicts the angle $\alpha_j$ at which the droplet is jetted out of the cavity. Not every impact leads to the observation of a jetted droplet, which could mean that no droplets were pinched off from the jet, or that droplets remain trapped inside the cavity. Given the small droplet size and the limited depth of field of our imaging system, the ejected droplet could also have moved out of the optical focal plane. 

Another observation is that the trajectory of the droplet is curved when it moves out of the cavity, as shown in Fig.~\ref{fig:trajectory}. This may be an indication that the initial $\alpha_j$, directly after pinch-off from the jet, is different from the $\alpha_j$ measured above the closing cavity, where it may be hindered by a disturbing airflow. To better quantify the jet direction, we therefore change our focus to the cavity collapse and deduce the retraction angle $\alpha_r$, as depicted in Fig.~\ref{fig:alphar}, where we assume that the onset of retraction of the cavity sets the initial direction of the jetted drop. From a series of frames $\alpha_r$ is measured where the first frame was taken when the cavity is at its maximum size. In the next frames we follow the direction of cavity collapse by tracking this point to find the retraction angle.

 \begin{figure}
\centering
  \begin{subfigure}[c] {0.60\textwidth}
	\includegraphics[width=\textwidth]{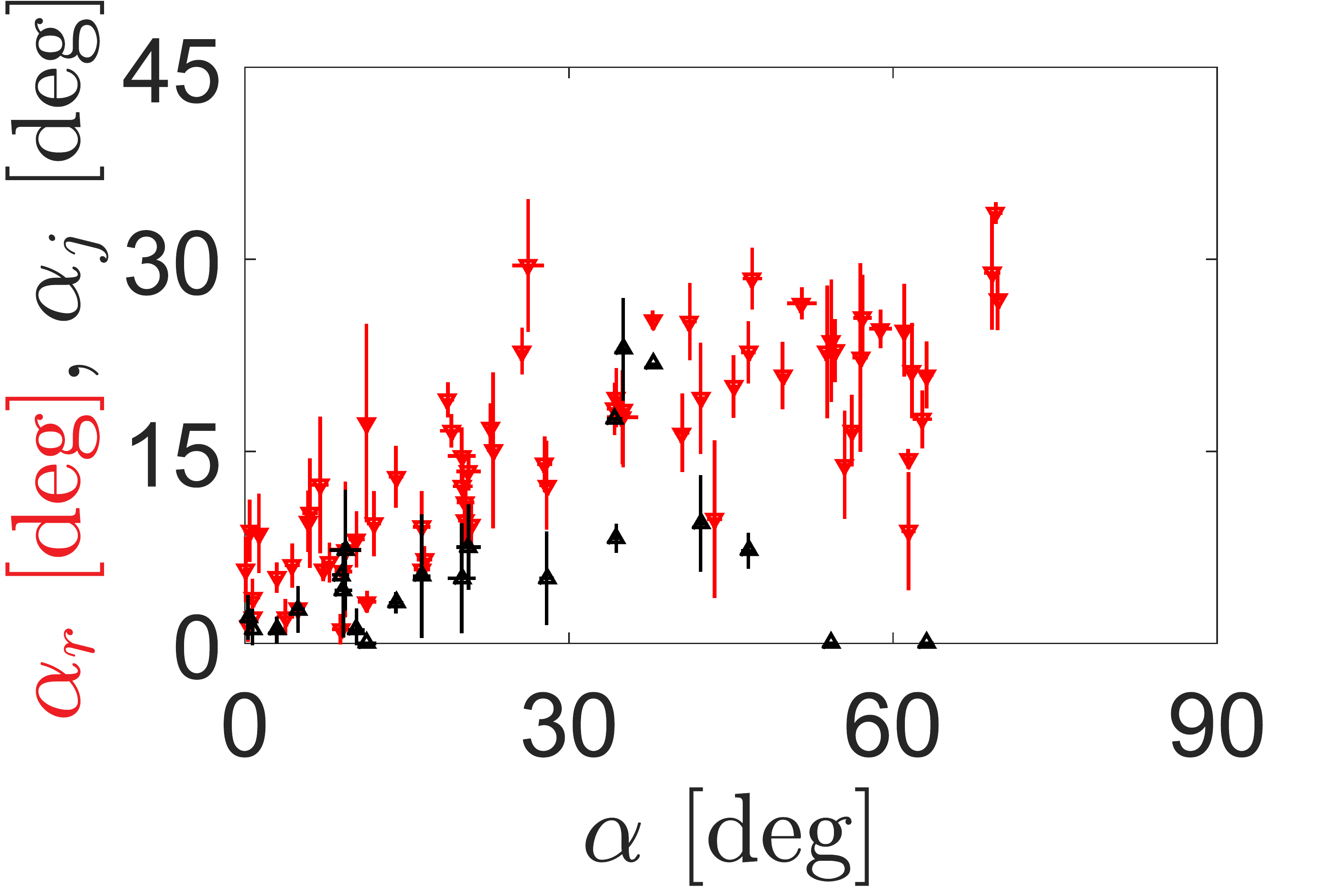} 
       \caption{}
       \label{fig:alphairj}
    \end{subfigure}
    \begin{subfigure}[c] {0.35\textwidth}
	\includegraphics[width=\textwidth]{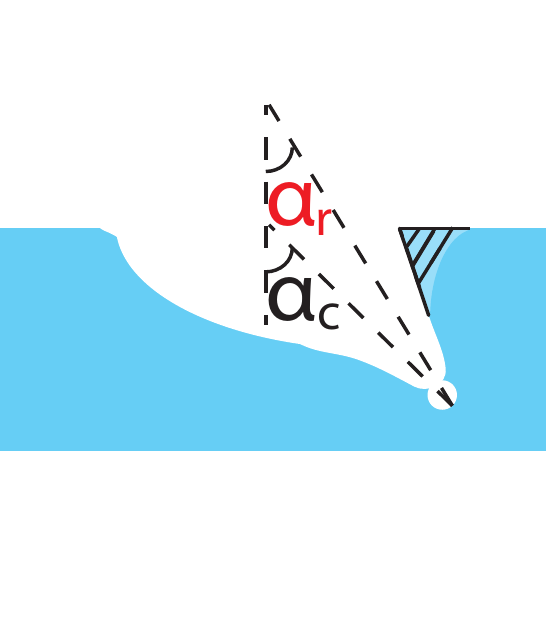} 
        \caption{}
        \label{fig:cavshift}
    \end{subfigure}
\caption{a)  Angle of the droplet jetted out of the cavity $\alpha_j$ (upward black triangles, each triangle corresponds to the average of about six measurements) and retraction angle of the cavity $\alpha_r$ (downward red triangles, each triangle corresponds to the average taken over six measurements below and above the water surface) as a function of the impact angle $\alpha$. b) Schematic view of the cavity angle and the retraction angle. When a drop impacts obliquely, one expects a cavity consisting of the white area, due to symmetry. However, since surface tension prohibits the existence of the sharp corner, the shaded area is pulled outward. Therefore, an asymmetric cavity forms resulting in a retraction angle which is smaller compared to the cavity angle.}
\label{fig:alpharcavshift}
\end{figure} 

As shown in Fig.~\ref{fig:alphairj}, $\alpha_j$ and $\alpha_r$ are indeed similar, with $\alpha_j$ being slightly smaller than $\alpha_r$, which is probably due to the influence of a disturbing airflow, as discussed above. $\alpha_r$ follows $\alpha$ up to about 25$^\circ$. For $\alpha >$ 25$^\circ$, $\alpha_r$ flattens and reaches a maximum of about 30$^\circ$. The deviation from $\alpha$ occurs around the same $\alpha$ for both $\alpha_c$ and $\alpha_r$. However, the build-up of drag waves in front of the impacting drop is not enough to account for the flattening observed for $\alpha_r$.
We speculate that this saturation is caused by a maximum angle that the cavity can make with the surface. For perpendicular impact, the cavity is symmetric and hemispherical, which is energetically most favorable. For oblique impact, the cavity is no longer hemispherical but one side of the cavity wall starts to move inwards (see Fig.~\ref{fig:cavshift}). At the other side the cavity opening grows as surface tension smooths the sharp edge as is illustrated in Fig.~\ref{fig:cavshift} by the shaded area. This effect breaks the symmetry and therefore poses a limit on the maximum value of $\alpha_r$.


\section{Discussion and Conclusions}
We presented an experimental study of oblique drop impact onto a quiescent deep liquid pool. We performed quantitative experiments where drops impact obliquely onto a deep liquid pool for a wide range of Weber numbers and impact angles and analyzed the splashing behavior, the cavity formation, and the cavity collapse.

In analogy to previous studies, e.g. \cite{zaleskisplash, Stow419, Sommerfeld1997, Josserand2016365}, we found that the crown velocity has to be larger than the Taylor-Culick velocity to obtain splashing. For oblique drop impact this crown velocity is influenced by the tangential velocity of the impacting drop, leading to an asymmetry in the crown and giving rise to an asymmetry in the splashing threshold~\cite{1367-2630-11-6-063017}. In contrast to the model presented in~\cite{1367-2630-11-6-063017} where oblique drop impact onto a dry substrate was studied, we cannot describe our data by simply adding/subtracting the tangential velocity to the crown velocity. Instead, we assumed the tangential velocity influences the amount of mass squeezed into the crown. We quantified this effect using scaling arguments and derived a model that is consistent with our measurements for a wide range of $We$ and $\alpha$. Thus, the model gives valuable insight into the occurrence and direction of splashing.
 
For small impact angle $\alpha$ the cavity angle $\alpha_c$ directly equals the impact angle and does not depend on $We$, for large $\alpha$ a decrease in $\alpha_c$ with respect to $\alpha$ was observed. The magnitude of the cavity displacement depends on the tangential Weber number and the cavity depth depends on the perpendicular Weber number. The scarce numerical simulations on oblique impact~\cite{Ray20121386} confirm this $We$-dependence of the cavity displacement. However, since~\cite{Ray20121386} described drop impact onto a thin liquid film, no information on the cavity depth is available. 

In conclusion, our study provides the first quantitative overview of the events following oblique impact onto a deep quiescent liquid pool. The results allow to predict under what impact velocity and in what direction drops can splash after impact. In our experiment, data from a single measurement plane is obtained. It would be interesting to obtain the full 3D profile above and below the water surface using holographic microscopy~\cite{Schnars2002} or by extended numerical simulations. Further studies will also be required to assess the influence of gravity on splashing and the cavity collapse, which may be relevant for larger droplets as generated by rain or breaking waves. 
\begin{acknowledgments}
We acknowledge the Industrial Partnership Programme of the Netherlands Organization for Scientific Research (NWO), and NanoNextNL, a micro and nanotechnology consortium of the Government of the Netherlands and 130 partners. We also acknowledge P.E. Frommhold and R. Mettin for providing the drop separator and K. Winkels, J.F. Dijksman and B. de Smet for stimulating discussions. 
\end{acknowledgments}

MVG and PS contributed equally to this work. 


%

\end{document}